\documentclass[twocolumn,prl,superscriptaddress,showpacs,amsmath,amssymb]{revtex4}
\usepackage{graphicx}
\usepackage{dcolumn}
\usepackage{bm}

\begin{document}
\title{Giant electro-thermal conductivity and spin-phonon coupling
in an antiferromagnetic oxide}
\author{C.~Chiorescu}
\affiliation{Department of Physics, University of Miami, Coral
Gables, Florida 33124}
\author{J.~J.~Neumeier}
\affiliation{Department of Physics, Montana State University,
Bozeman, Montana 59717}
\author{J.~L.~Cohn}
\affiliation{Department of Physics, University of Miami, Coral
Gables, Florida 33124}

\begin{abstract}
The application of weak electric fields ($\lesssim 100$~V/cm) is
found to dramatically enhance the lattice thermal conductivity of
the antiferromagnetic (AF) insulator CaMnO$_3$ over a broad range
of temperature about the N\'eel ordering point (125 K). The effect
is coincident with field-induced de-trapping of bound electrons,
suggesting that phonon scattering associated with short- and
long-ranged AF order is suppressed in the presence of the
mobilized charge. This interplay between bound charge and
spin-phonon coupling might allow for the reversible control
of spin fluctuations using weak external fields.

\end{abstract}

\pacs{66.70.-f,75.47.Lx, 75.80.+q, 71.55.-i}
\maketitle

The strong magneto-elastic coupling present in certain
antiferromagnetic (AF) ferroelectrics \cite{GiantME}
(e.g.,~hexagonal YMnO$_3$) has attracted considerable attention
recently because of its importance in mediating a coupling between
magnetic and electric orders \cite{NatureReview} and the potential
of such multiferroics for applications \cite{Apps}.  The hexagonal
manganites also possess very strong spin fluctuations
characteristic of a spin liquid well above their N\'eel
temperatures, possibly related to geometric spin frustration. This
raises the prospect of manipulating {\it dynamic} magneto-elastic
coupling in the paramagnetic phase (PM) of these or other
compounds using external fields.

Dramatic signatures of such coupling in YMnO$_3$ are a suppressed
thermal conductivity \cite{Sharma} ($\kappa$) and anomalous
ultrasonic attenuation \cite{Poirier} over a broad temperature
range in the PM phase. Two AF oxides with quite similar features
in their thermal conductivities are the transition-metal monoxide,
MnO \cite{SlackMnO} ($T_N=118$~K), and orthorhombic CaMnO$_3$
\cite{ZhouGoodenough,CohnNeumeier} ($T_N\sim 125$~K). Like
YMnO$_3$, both of these materials possess large exchange striction
\cite{MnOStriction,CMOStriction} (with changes in lattice
constants at $T<T_N$, $\Delta a/a\gtrsim 10^{-4}$), significant
spin frustration \cite{MnOSpinFluc,NeumeierGoodwin}
$\Theta/T_N\sim 4-5$ ($\Theta$ is the Curie-Weiss temperature),
and strong short-range spin correlations extending well above
$T_N$ \cite{MnOSpinFluc,Granado}. The suppressed thermal
conductivities in the PM phase of YMnO$_3$ and CaMnO$_3$ (and by
extension, MnO) have been proposed to arise from the scattering of
acoustic phonons by nanoscale strains generated by short-ranged
spin correlations \cite{Sharma,ZhouGoodenough}.  This scattering
diminishes rapidly when long range order is established, giving
rise to sharp increases in $\kappa$ at $T<T_N$.

CaMnO$_3$ is distinguished from YMnO$_3$ and MnO by the presence
of low-lying electron donor levels (oxygen vacancies) from which a
small density of mobile electrons ($\Delta n\sim
10^{16}$~cm$^{-3}$) can be released under relatively weak applied
electric fields ($F\lesssim 100$~V/cm) \cite{PriorHall}. Here we
demonstrate that these mobilized carriers are associated with a
substantial reduction in phonon scattering, yielding relative
changes in $\kappa$ with field, $(1/\kappa)(\Delta \kappa/\Delta
F)$ that are two orders of magnitude larger than found in quantum
paraelectrics such as KTaO$_3$ and SrTiO$_3$ \cite{Goldman} where
the field couples to soft-mode phonons.  The apparent absence of
permanent electric dipoles or lattice anomalies in CaMnO$_3$ leads
to the hypothesis that the mobilized electrons themselves mediate
a suppression of phonon scattering from strain induced by short-
and long-ranged magnetic order.

Measurements were performed on single-crystal and polycrystalline
specimens of CaMnO$_3$.  Their preparation methods and physical properties
have been reported elsewhere \cite{CohnNeumeier,PriorHall,NeumeierCohn,CohnEps,CohnPolarons}.
We focus here on the crystal (dimensions $2\times0.9\times0.05\
{\rm mm}^3$) for which $\kappa$ was measured for two different
oxygen configurations produced by annealing at 600$^{\circ}$~C in
flowing oxygen and air, respectively. The net carrier density for
the oxygen-annealed state, as determined from room-temperature
Hall measurements \cite{PriorHall}, was $N\simeq 6\times
10^{18}$~cm$^{-3}$. This corresponds to $\sim 3\times 10^{-4}$
electrons per formula unit, attributed to a very small oxygen
deficiency. The carrier density after annealing in air is
estimated \cite{CohnPolarons} from the increase in
room-temperature thermopower (-530~$\mu$/K to -370 $\mu$/K) to be
$N\simeq 4\times 10^{19}$~cm$^{-3}$.  $\kappa$ was measured with a
steady-state technique employing a heater and 25 $\mu$m-diameter
chromel-constantan thermocouple; four-probe electrical resistivity
($\rho$) was measured during the same experiments.  $\kappa(I,T)$
was measured for the air-annealed crystal in the presence of {\it
dc} transport currents ($I=5$~nA--10 mA).  With the specimen
suspended in vacuum and thermally anchored at only one end, Joule
heating was significant at the highest currents. The average
specimen temperature (relative to a resistance sensor mounted on
the cold stage) was monitored with a second thermocouple following
application of a transport current, and allowed to stabilize prior
to energizing the specimen heater for determining $\kappa$.
Linearity in the heater power - $\Delta T$ response was confirmed
at various temperatures throughout the measurement range.

\begin{figure}
\includegraphics[width = 3.25in,clip]{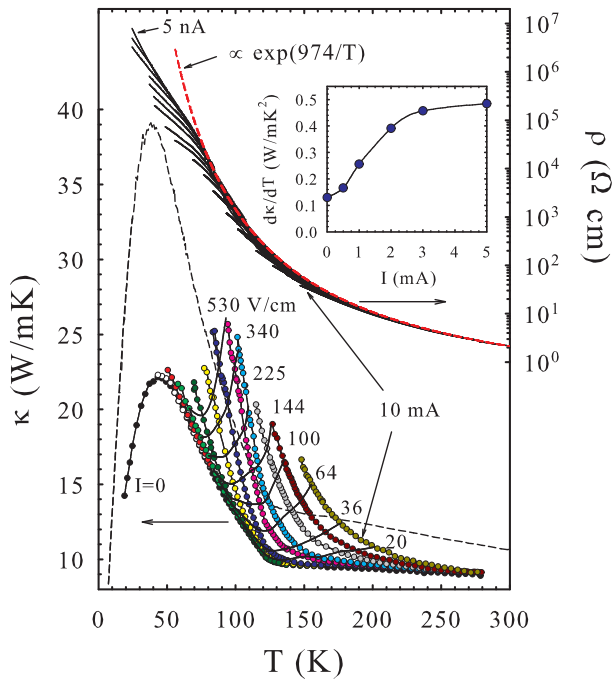}%
\caption{(Color online) Lower panel (left ordinate): $\kappa(T)$
for the air-annealed CaMnO$_3$ crystal (circles) at various
transport currents, from right to left (in mA): 10, 5, 3, 2, 1,
0.5, 0.25, 0.1, 0.05, 0.01, 0. The solid curves bridging
$\kappa(T)$ plots are contours of constant electric field (values
labeled).  The dashed curve represents data for the
oxygen-annealed crystal for $I=0$. Upper panel (right ordinate):
$\rho(T)$ for the same currents as $\kappa(T)$ as well as the
following (in $\mu$A): 5, 3, 1, 0.3, 0.1, 0.03, 0.01, 0.005.
Inset: current dependence of the slope, $d\kappa/dT$, for air
annealing.}
\label{KofIT}
\vglue -.1in
\end{figure}

Figure~\ref{KofIT} (lower panel) shows $\kappa(T)$ after oxygen
(dashed curve) and air (circles) annealing.  The $I=0$ and
$I=10$~mA plots for air annealing are labeled.  Constant electric
field ($F$) contours (with values labeled) are represented by
solid curves bridging the constant current plots.  These fields,
$F=\rho J$, were determined from $\rho(T)$ data (upper panel). The
abrupt upturn in $\kappa$ upon entering the AF ordered phase is
evident near $T_N\approx 120$~K.  Note that for the $I=0$ curves
the slope just below $T_N$, $d\kappa/dT|_{T_N}$, is greater for oxygen
annealing, and increases with $I$ for the air annealed crystal
(inset, Fig.~\ref{KofIT}).  These features are discussed further below.
The data were reproducible upon thermal and current cycling, and after
subsequent re-annealing.

\begin{figure}
\includegraphics[width = 2.8in,clip]{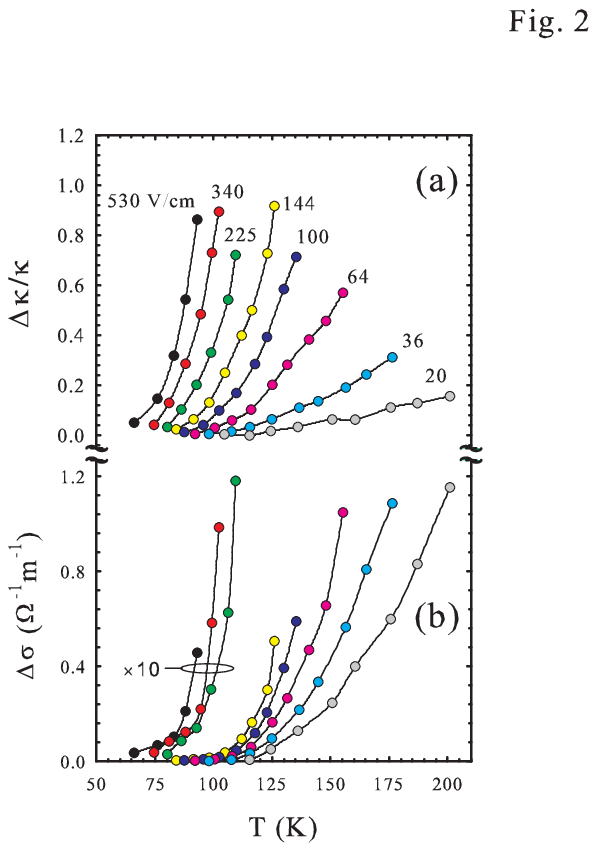}%
\caption{(Color online) (a) Fractional change in thermal
conductivity {\it vs} temperature at constant electric fields, (b)
Field-induced change in electrical conductivity for the same field
values as in (a).}
\label{DeltaK}
\vglue -.2in
\end{figure}

Figure~\ref{DeltaK}~(a) shows the $T$-dependence of the fractional
change in thermal conductivity,
$(\Delta\kappa/\kappa)\equiv[\kappa(F,T)-\kappa(0,T)]/\kappa(0,T)$,
computed for these values of $F$.  The enhancement of $\kappa$ in
field is substantial, approaching 100\% for fields $F>100$~V/cm in
the range $80$~K~$<T<130$~K, and becomes negligible at lower
temperatures. Qualitatively similar results were found for the
polycrystalline specimen, with a temperature interval for the
largest field effects lower by about 50 K.

The field-enhanced $\kappa$ correlates with
an increased electrical conductivity,
$\Delta\sigma=\sigma(F,T)-\sigma(0,T)$ [Fig.~\ref{DeltaK}~(b)].
As demonstrated by prior Hall effect studies on this same
crystal \cite{PriorHall}, the mechanism underlying the increase in
$\sigma$ is Poole-Frenkel \cite{PooleFrenkel} release of carriers from
Coulomb traps (oxygen vacancy donor levels) via field-assisted barrier-lowering.
However, it is clear that the thermal conductivity of these mobilized electrons
cannot account for the increase in $\kappa$. For example,
$\Delta\sigma\sim 0.5\ \Omega^{-1}{\rm m}^{-1}$ at $F=144$~V/cm and $T=126$~K,
corresponding to an increase in the
electron density, $\Delta n\simeq 3\times 10^{16}$~cm$^{-3}$
(using a mobility \cite{CohnPolarons} $\sim 1$~cm$^2$/V-s), about $50$\%
of the equilibrium value at that temperature.
The Wiedemann-Franz law, which provides an upper-bound estimate on the
electronic thermal conductivity, yields
$\Delta\kappa_e=L_0\Delta\sigma T\simeq 1.5\times10^{-6}$~W/mK,
nearly seven orders of magnitude smaller than observed.  Spin-wave
heat conduction, though a possible contributor to $\kappa$ at low
$T$ in the ordered phase, cannot account for $\kappa$ well above $T_N$.
Thus the increase of $\kappa$ must have its origin in reduced phonon scattering.

Bound electrons may cause phonon scattering through static local
distortions of bond lengths \cite{CohnKLCMO} or more extended,
dynamic polaronic distortions, and possibly the diminution of this
scattering as electrons become mobile in field explains the
observations.  But this proposal must be viewed as implausible
given that the fraction of bound electrons released is very small,
$\sim \Delta n/N\lesssim 10^{-3}$, so that the density of such
distortions is altered negligibly in field. Further supporting
this view, $\kappa$ at the highest $F$ for the air-annealed
crystal exceeds that in zero-field for the oxygen-annealed state
(Fig.~\ref{KofIT}), in spite of the fact that the bound electron
density in the former exceeds that of the latter by an order of
magnitude.  In addition, there is no evidence of permanent
electric dipoles in this material \cite{CohnEps,Spaldin}, the
alignment of which in weak fields might conceivably alter the
phonon scattering.  This leads us to consider that the {\it
mobilized} electrons themselves mediate a suppression of phonon
scattering.

In order to assess how this might occur and to gain insight into
the origin of the strong PM-phase phonon scattering, it is
instructive to examine changes in structure induced by the onset
of magnetic order. Anomalous Mn motion, stabilized at $T<T_N$, was
recently proposed to explain the $\kappa$ behavior of hexagonal
YMnO$_3$ \cite{ZhouSoftModes}. However, the unusual Mn-O
bondlength changes \cite{GiantME} motivating this proposal are
absent in the average and local structure of CaMnO$_3$
\cite{Bozin}. On the other hand, rotations of the MnO$_6$
octahedra couple strongly to the magnetic order in CaMnO$_3$
\cite{Granado,Bozin} and provide a mechanism through which spin
fluctuations at $T>T_N$ may induce strain that scatters
heat-carrying acoustic modes. Perturbations due to the polaronic
motion \cite{PriorHall,CohnPolarons,SpinPolarons} of mobilized
electrons may compete with the distortions favored by short-ranged
AF ordered domains.

Consider the scenario for phonon scattering by spin fluctuations
introduced by Sharma {\it et al.} \cite{Sharma}. Strain fields
associated with short-ranged AF spin-ordered regions have spatial
extent given by the spin correlation length $\xi$.  Slowly
fluctuating on the timescale of lattice vibrations, they scatter
acoustic phonons. For the case $\lambda_p\ll \xi$ (where
$\lambda_p$ is the phonon wavelength \cite{NoteOnLengths}), their
scattering rate can be approximated (for a spherical
spin-correlated domain) as, $\tau_{mag}^{-1}=p_0v\pi(\xi/2)^2$
($p_0$ and $v$ are the density of scatterers and phonon velocity,
respectively). This scattering should be sharply diminished at
$T<T_N$ as long-range magnetic order is established.  Residual
strain in AF domain walls and other defects (e.g. twins) likely
scatter phonons to low temperatures as evidenced by a phonon
mean-free path \cite{MFP}, $\ell_{ph}=3\kappa/Cv$, that remains $<
1\mu$m at the lowest measured $T$.

To examine this picture more quantitatively we employ the
Debye-Calloway model \cite{Berman} to compute the lattice thermal
conductivity, adding $\tau_{mag}^{-1}$ to a sum of traditional
phonon scattering rates arising from other phonons (Umklapp),
dislocations, and point-like defects \cite{CallowayFits}. We
restrict our analysis to the PM phase and for simplicity ignore
possible temperature dependencies of $n_0$ and $\xi$, taking
$\gamma\equiv p_0\pi(\xi/2)^2$ as a constant parameter for
$T>T_N$.  The $I=0$ data for air and oxygen annealing were first
fitted by assuming the traditional phonon scattering terms to be
independent of the annealing treatment, and the computed $\kappa$
in the absence of magnetic scattering ($\gamma=0$) was constrained
to match the oxygen annealed data at the lowest $T$ and to exceed
the highest measured values (including those for air annealing in
applied field) at intermediate $T$.  Subsequently, the data for
air annealing in applied field were fitted with $\gamma$ as the
only adjustable parameter.

\begin{figure}
\includegraphics[width = 3.2in,clip]{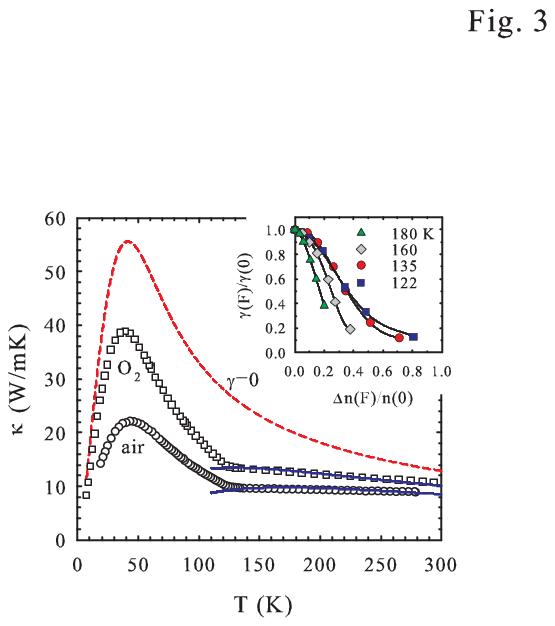}%
\caption{(Color online) Comparison of calculated (curves) and
measured (symbols) $\kappa(T)$ for $I=0$. The dashed curve is the
computed $\kappa$ in the absence of magnetic scattering
($\gamma=0$). The solid curves used $\gamma=2.51 (0.82)\times
10^7{\rm m}^{-1}$ for air (oxygen) annealing. Inset: magnetic
scattering strength versus field-induced carrier density
determined from fits to $\kappa(F)$ at fixed $T$.} \label{Calcs}
\end{figure}
Fig.~\ref{Calcs} compares the calculated (solid curves) and measured $\kappa(T)$.
Also shown is the hypothetical $\kappa(T)$ in the absence of
magnetic scattering ($\gamma=0$, dashed curve).  The values of $\gamma=2.51
(0.82) \times 10^7$~m$^{-1}$ for air (oxygen) annealing are similar to that employed for
YMnO$_3$ \cite{Sharma}.  For self-consistency, the distance between scatterers
$d=2(3/4\pi n_0)^{1/3}$, must be larger than their size $\xi$.
Both YMnO$_3$ \cite{Sharma} and MnO \cite{MnOSpinFluc} have
comparable values of $\xi\sim 20-50{\rm\AA}$.  In the absence of a
direct measurement of $\xi$ for CaMnO$_3$, taking $\xi=40\rm{\AA}$
implies a density of scatterers $p_0=2.0 \times
10^{18}$~cm$^{-3}$ for air annealing, or equivalently,
$d\simeq 100{\rm\AA}>\xi$ as required.

Fitting the model to $\kappa(F)$ at fixed $T$ yields the decrease
in magnetic scattering {\it vs.} the relative change in mobile
carrier density (inset of Fig.~\ref{Calcs}). The decrease in
$\gamma$, by up to a factor of 10 at the highest applied fields
near $T_N$, implies a decrease in $p_0$, $\xi$ or both.

The nucleation of AF domains in the PM phase and their growth in
the ordered phase likely depend on domain-wall pinning near
lattice defects \cite{Pinning}. Mobilized electrons could reduce
$p_0$ by depinning AF domains, an idea motivated by comparing the
zero-field data for air and oxygen annealing. The larger $\kappa$
and $d\kappa/dT|_{T_N}$ observed for the more oxygenated crystal
in the ordered phase (Fig.~\ref{KofIT}) suggests a coarsening of
the domain structure, consistent with observations in
Gd$_2$(MoO$_4$)$_3$ of strong phonon scattering at ferroelastic
domain boundaries \cite{Ferroelastic}. Oxygen vacancies or their
complexes are thus implicated in pinning (the estimate of $p_0$
above suggests that $\sim 10\%$ of vacancies participate in
pinning).

The correlation between $\kappa$ and $d\kappa/dT|_{T_N}$ in
applied electric field (Fig.~\ref{KofIT}) suggests that mobilized
electrons cause depinning.  The polaronic motion of charge
carriers can increase the frequency of spin fluctuations
\cite{214}, effectively reducing $p_0$. Regarding $\xi$ there is
precedent for its diminution by light charge-carrier doping in
Ca$_{1-x}$La$_x$MnO$_3$ \cite{Granado} and La$_{2-x}$Sr$_x$CuO$_4$
\cite{214}. However, the carrier densities in these cases are
several orders of magnitude larger than $\Delta n$ induced by
field in the present experiments. Both $\kappa$ and
$d\kappa/dT|_{T_N}$ decrease with substitutional doping in
Ca$_{1-x}$La$_x$MnO$_3$ \cite{CohnNeumeier}, contrary to the
field-induced trend, but this may simply reflect the predominance
of additional scattering from substitutional disorder. More direct
measurements of the AF correlations in CaMnO$_3$ as functions of
temperature and applied electric field are clearly called for to
further test and refine the ideas put forward here.

In summary, the present measurements of thermal conductivity in
CaMnO$_3$ imply that a small density of electrons released from
donor states in weak electric fields substantially reduce the
scattering of phonons associated with short- and long-ranged AF
order. The results suggest that the correlated spin volume in the
PM phase is highly sensitive to the injected carrier density and
reversibly controlled by a transport current. Strong AF spin
fluctuations, magnetostriction, and the presence of loosely-bound
charges in defect states appear to be key ingredients underlying
the observations. This novel manipulation of dynamic
magneto-elastic coupling motivates further studies of spin
correlations in this compound or others having similar
characteristics.

This material is based upon work supported by the National Science
Foundation under grants DMR-0072276 (Univ.~Miami) and DMR-0504769
(Mont.~St.~Univ.), and the Research Corporation (Univ.~Miami).

\newpage

\begin{thebibliography}{99}

\bibitem{GiantME} S. Lee {\it et al.}, Nature {\bf 451}, 805 (2008);
J. Cao {\it et al.}, Phys. Rev. Lett. {\bf 100}, 177205 (2008);
D. Meier {\it et al.}, New. J. Phys. {\bf 9}, 100 (2007); F. Ye {\it et al.},
Phys. Rev. B {\bf 73}, 220404(R) (2006).

\bibitem{NatureReview} W. Eerenstein, N. D. Mathur, and J. F. Scott, Nature {\bf 442}, 759 (2006).

\bibitem{Apps} M. Fiebig, J. Phys. D: Appl. Phys {\bf 38}, R123 (2005);
T. Kimura {\it  et al.}, Nature {\bf 426}, 55 (2003); N. Hur {\it  et al.},
{\it ibid.} {\bf 429}, 392 (2004); T. Lottermoser {\it  et al.}, {\it ibid.} {\bf 430}, 541 (2004).

\bibitem{Sharma} P. A. Sharma {\it et al.} Phys. Rev. lett. {\bf 93}, 177202 (2004).

\bibitem{Poirier} M. Poirier {\it et al.}, Phys. Rev. B {\bf 76}, 174426 (2007).

\bibitem{SlackMnO} G. A. Slack and R. Newman, Phys. Rev. Lett. {\bf 1}, 359 (1958).

\bibitem{ZhouGoodenough} J.-S. Zhou and J. B. Goodenough, Phys. Rev. B {\bf 66}, 052401 (2002).

\bibitem{CohnNeumeier} J. L. Cohn and J. J. Neumeier, Phys. Rev. B {\bf 66}, 100404(R) (2002).

\bibitem{MnOStriction} B. Morosin, Phys. Rev. B {\bf 1}, 236 (1970).

\bibitem{CMOStriction} Y. Moritomo {\it et al.}, Phys. Rev. B {\bf 64}, 214409 (2001).

\bibitem{MnOSpinFluc} A. Renninger, S. C. Moss, and B. L. Averbach, Phys. Rev. {\bf 147}, 418
(1966); H. Betsuyaku, Sol. St. Commun. {\bf 26}, 345 (1977).

\bibitem{NeumeierGoodwin} J. J. Neumeier and D. H. Goodwin, J. Appl. Phys. {\bf 85}, 5591 (1999).

\bibitem{Granado} E. Granado {\it et al.}, Phys. Rev. Lett. {\bf 86}, 5385 (2001).

\bibitem{PriorHall} C. Chiorescu, J. L. Cohn, and J. J. Neumeier, B {\bf 76}, 020404(R) (2007).

\bibitem{Goldman} W. H. Huber, L. M. Hernandez, and A. M. Goldman, Phys. Rev. B {\bf 62}, 8588
(2000).

\bibitem{NeumeierCohn} J. J. Neumeier and J. L. Cohn, Phys. Rev. B {\bf 61} 14319 (2000).

\bibitem{CohnEps} J. L. Cohn, M. Peterca, and J. J. Neumeier. Phys. Rev. B {\bf 70} 214433 (2004).

\bibitem{CohnPolarons} J. L. Cohn, C. Chiorescu, and J. J. Neumeier, Phys. Rev. B {\bf 72}, 024422 (2005);
C. Chiorescu, J. J. Neumeier, and J. L. Cohn, {\it ibid.} {\bf 73}, 014406 (2006).

\bibitem{Chmaissem} O. Chmaissem {\it et al.}, Phys. Rev. B {\bf 64}, 134412 (2001).

\bibitem{PooleFrenkel} J. Frenkel, Phys. Rev. {\bf 54}, 647 (1938).

\bibitem{CohnKLCMO} J. L. Cohn {\it et al.}, Phys. Rev. B {\bf 56}, R8495 (1997).

\bibitem{Spaldin} A. Filippetti and N. A. Hill, Phys. Rev. B {\bf 65}, 195120 (2002).

\bibitem{ZhouSoftModes} J.-S. Zhou {\it et al.}, Phys. Rev. B {\bf 74}, 014422 (2006).

\bibitem{Bozin} E. S. Bozin {\it et al.}, J. Phys. Chem. Sol. {\bf 69}, 2146 (2008).

\bibitem{SpinPolarons} Y. R. Chen and P. B. Allen, Phys. Rev. B {\bf 64}, 064401 (2001);
H. Meskine, T. Saha-Dasgupta, and S. Satpathy, Phys. Rev. Lett.
{\bf 92}, 056401 (2004); H. Meskine and S. Satpathy, J. Phys.:
Condens. Matter {\bf 17}, 1889 (2005).

\bibitem{NoteOnLengths} A resonable estimate for thermal phonons is
$\lambda_p\sim hv/(3.8k_BT)\approx 2-6{\rm\AA}$ in the PM phase.

\bibitem{MFP} Specific heat ($C$) data for CaMnO$_3$ are reported by
Y. Moritomo {\it et al.}, Phys. Rev. B {\bf 61}, 204409  (2001)
and A. Cornelius {\it et al.}, {\it ibid.} {\bf 68}, 014403
(2003). The sound velocity $v=4,800$~m/s employed represents an
average computed from the experimental Debye temperature
($\Theta_D$).

\bibitem{Berman} R. Berman, {\it Thermal Conduction in Solids} (Clarendon, Oxford, 1976).

\bibitem{CallowayFits} The scattering of phonons of frequency $\omega=vq$ by other phonons
(U-processes), dislocations, and point defects were
represented by rates: $A_1\omega^2T\exp(-\Theta_D/bT)$,
$A_2\omega$, $A_3\omega^4$.  Boundary scattering was found to be negligible in the measured
temperature range.  Parameter values (the same for air and oxygen annealing)
were: $A_1=2.63\times 10^{-18}$~s~K$^{-1}$ , $A_2=7.65\times 10^{-5}$,
and $A_3=2.36\times 10^{-43}$~s$^{3}$.  These values are comparable to
those found for YMnO$_3$ (Ref.~\onlinecite{Sharma}). $b$ was taken as
5 (Ref.~\onlinecite{Berman}).

\bibitem{Pinning} Yin-Yuan Li, Phys. Rev. {\bf 101}, 1450 (1956);
B. K. Tanner, Contemp. Phys. {\bf 20}, 187 (1979).

\bibitem{Ferroelastic} S. Mielcarek {\it et al.}, Proceedings of the 11th IEEE
Int'l Symposium on Applications of Ferroelectrics, ISAF98, p. 415 (1998).

\bibitem{214} V. Kataev {\it et al.}, J. Phys.: Condens. Matter {\bf 11}, 6571 (1999);
K. Hirota {\it et al.}, Physica C {357-360}, 61 (2001).

\end{thebibliography}
\end{document}